\documentclass[reprint,amssymb, amsmath, aip,cha,twocolumn]{revtex4-1}
\usepackage{graphicx}
\usepackage{amsmath}
\usepackage[caption = false]{subfig}
\usepackage{color}

\usepackage{epsfig}
\usepackage{bm}%
\usepackage[colorlinks=true,linkcolor=blue]{hyperref}%
\expandafter\ifx\csname package@font\endcsname\relax\else
 \expandafter\expandafter
 \expandafter\usepackage
 \expandafter\expandafter
 \expandafter{\csname package@font\endcsname}%
\fi
\begin{document}
\title{Micro-dynamics of neutral flow induced dusty plasma flow}
\author{Garima Arora}%
\email{garimagarora@gmail.com}
\author{P. Bandyopadhyay}
\author{M. G. Hariprasad}
\author{A. Sen}
\affiliation{Institute For Plasma Research, HBNI, Bhat, Gandhinagar,Gujarat, India, 382428}%
\date{\today}
\begin{abstract}
We present a detailed experimental study of gas flow induced motion of dust particles in a DC glow discharge plasma. The characteristics of the dust dynamics are investigated as a function of the differential gas flow rate, the background neutral pressure, the dust particle size as well as the neutral species of the gas. The experiments have been carried out in the table top Dusty Plasma Experimental (DPEx) device in which a plasma is created between a disk shaped anode and a grounded cathode in a $\Pi$-shaped pyrex glass tube. The asymptotic steady state flow velocity of the injected  micron sized dust particles is found to increase with an increase of neutral flow velocity and decrease with an increase in the background pressure. Furthermore, this velocity is seen to be independent of the size of the dust particles but decreases with an increase in the mass of the background gas. A simple theoretical model, based on estimates of the various forces acting on the dust particles, is used to elucidate the role of neutrals in the flow dynamics of the dust particles. Our experiments thus provide a detailed microscopic understanding of some of the past phenomenological observations of dust flows in the DPEx device and can prove useful in future experimental implementations of dust flow experiments.  
\end{abstract}
\maketitle
\section{Introduction}\label{sec:intro}
\label{sec:introduction}
Dusty plasmas, consisting of a suspension of highly charged micron or sub-micron sized particles in a conventional electron-ion plasma \cite{basicscomplexplasmamorfill,basiccomplexplasmashukla}, have been the subject of a very large number of theoretical and experimental studies over the past few decades on account of their novel collective properties and variety of applications ranging from astrophysics to space physics to laboratory scale devices and industrial processes \cite{dustyplasmasolarsystem,tokomaksheath,roth1985spatial,fortov1997crystalline,chu1994direct,merlino2006dusty}. In laboratory devices a stationary dust cloud is normally formed near the plasma sheath boundary where the electric field of the sheath 
helps levitate the charged dust particles against the downward force of the gravity. Lateral confinement is usually provided by radial forces from externally applied potential fields. The levitated dust particles are also subject to a number of other forces due to their interaction with the background plasma and neutral gas and these can influence their dynamical state. A number of past studies have investigated the effect of such secondary forces like the radiation force \cite{radiationforce1, radiationforce2} the ion \cite{kharpakiondragforce,kharpakiondragforce2,barnesiondragforce} and neutral drag forces \cite{thomas,vladmirov, hanlonstreaming, yaroshenko2005determination} and the thermopherotic force \cite{iondragthermo} on dust dynamics and have explored means of maneuvering the dust dynamics through these forces. The ion and neutral drag forces, in particular, can play a dominant role in either inducing or limiting the steady state flow velocities of dust particles as they stream in a background of ions and neutrals.\\

The study of flowing dusty plasmas has attracted some recent attention and flow induced excitations in dusty plasmas have become an emerging area of research in laboratory as well as in the astrophysical context. When the dust fluid flows past an obstacle it can excite a wide variety of linear and nonlinear waves such as wakes and precursor solitons  \cite{surbhiPrecursor, senprecursor, sanatprecursor}, bow shocks \cite{bow, merlino2012dusty}, dispersive shocks \cite{surbhishock} and Von K\'{a}rman vortex streets \cite{charan2016molecular}. In these experimental studies, the flow in the dust fluid was typically initiated either by varying the confining potential \cite{surbhiPrecursor, merlino2012dusty} or by sudden changes in the gas flow into the chamber \cite{surbhishock} or by gravitational force through  mechanical tilting of one side of the experimental chamber \cite{bow}. The focus of these experiments so far have mainly been on the collective excitations in the dusty plasma medium and the dynamics of the dust flow itself has not received much attention. An exception to this is the detailed study of the ion drag force carried out by Yaroshenko {\it et al.} \cite{yaroshenko2005determination} on the PK4 device  over a range of discharge parameters and using particles of different sizes. The objective of the present paper is to present a similar detailed experimental study of the role of neutral drag in the dynamics of dust flow. A primary motivation for this work is to provide a better understanding of the role of neutrals in the dynamics of gas flow induced dusty plasma flows that were used in past experiments on the DPEx device \cite{surbhishock}. Our experiments have therefore been carried out in the same device over a range of gas flow conditions, background neutral pressures, different species of background ions and different dust particle sizes in order to elucidate the details of the dust flow dynamics. Extensive visual images of the dust particle trajectories have been collected and analyzed to trace their trajectories as they start from rest and accelerate to finally reach an asymptotic steady state velocity. It is found that this steady state velocity is strongly dependent on the gas flow and background neutral pressure - it increases with an increase in the flow velocity and decreases with an increase in the neutral pressure. It is also observed that although lighter dust particles accelerate faster the final steady state velocity value is independent of the size of the particles when all other conditions are kept the same. The size of the neutral particles however directly impacts the velocity - it decreases with an increase in the mass of the neutral species. Theoretical estimates of the various forces acting on the dust particles reveal that the neutral streaming force is the dominant one in inducing the flow. A simple model equation based on such a force is used to provide a qualitative description of the dust motion and a physical understanding of the dynamics.\\
  
The paper is organized as follows: In the next section (See.~\ref{sec:setup}), a brief description of the experimental setup along with the production of plasma and dusty plasma is presented. The generation of flow in dust particles and their velocity measurement are discussed in Sec. \ref{sec:flow}. The experimental observations, results and a simple model to describe the dust particle motion are discussed in Sec.~\ref{sec:results}. Some final concluding remarks are made in Sec. \ref{sec:conclusions}.   
\section{Experimental Set-up}\label{sec:setup}
\begin{figure}[ht]
\centering{\includegraphics[scale=0.47]{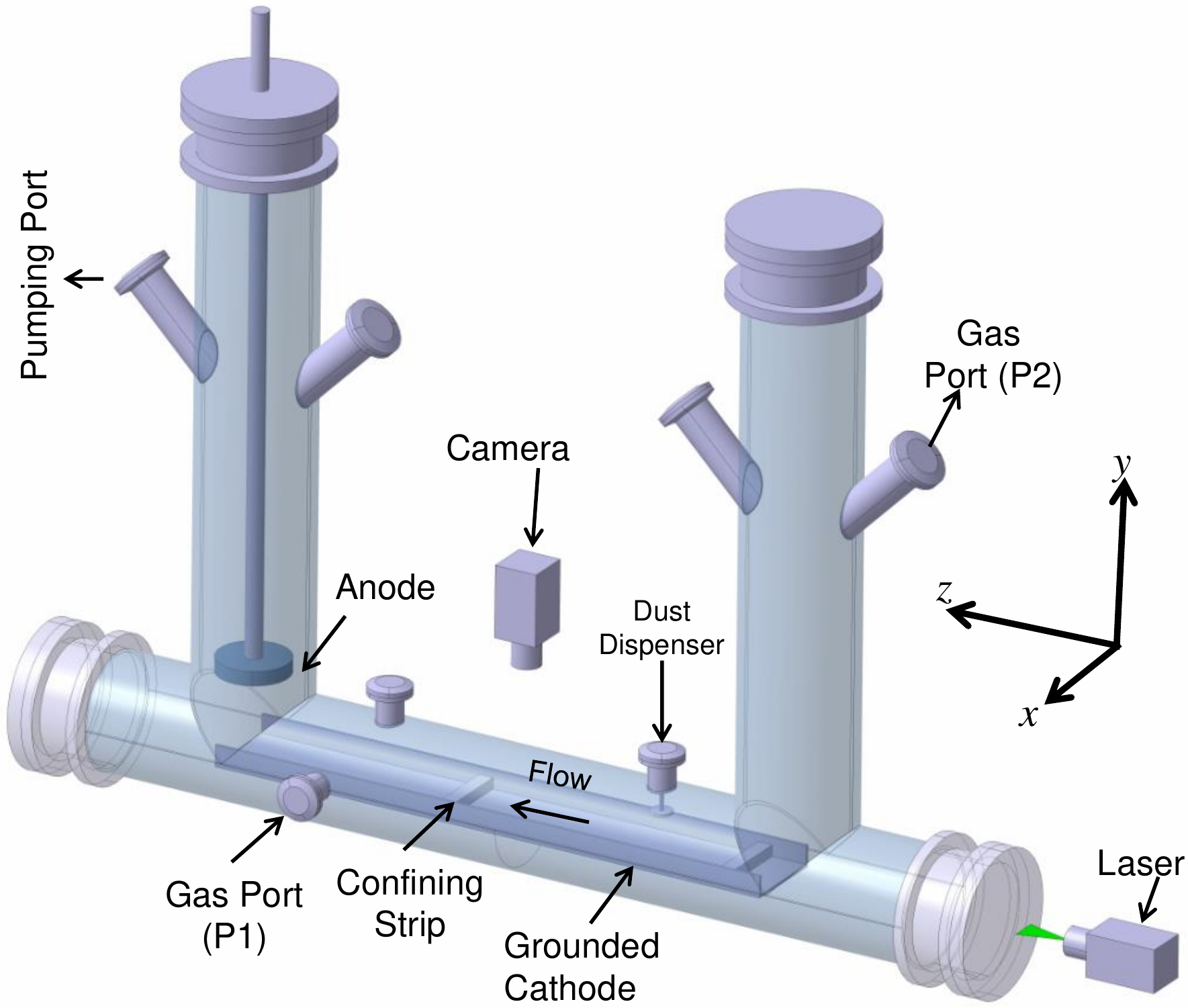}}
\caption{\label{fig:fig1} A schematic of the experimental set up for performing flowing dusty plasma experiments. }
\end{figure}
Our present set of experiments were carried out in the $\Pi$ shaped Dusty Plasma Experimental (DPEx) device. The system geometry of this device and the associated diagnostics have been given in details before \cite{surbhirsi}. A stainless steel (SS) circular disc of diameter 6 cm is hung from one arm of the $\Pi$-shaped tube, which acts as a live anode and a long SS plate tray of 40 cm length, 6 cm width and 2 mm thickness is used as a  grounded cathode for the purpose of plasma production. The sides of the cathode are bent to provide radial confinement of the dust particles and an additional two strips (6 cm $\times$ 1.2 cm $\times$ 0.5 cm) are placed on the cathode at a distance of 10~cm for axial confinement. The vacuum chamber is evacuated to a base pressure of 0.1 Pa with the help of a rotary pump. Argon gas is then introduced by a mass flow controller and/or by a gas dosing valve using either port-P1 and/or port-P2 depending on the experimental requirements to set the working pressure at $9-15$ Pa. A Direct Current (DC) voltage ($300-400$ V) is applied between the anode and the grounded cathode to produce plasma from the ambient Argon gas. The discharge current varies from $2-5$ mA. The plasma parameters such as the ion density ($n_i$) $\sim 1-3$ $\times 10^{15}/m^{3}$, the electron temperature ($T_e$) $\sim$ $2-4$ eV are measured using a single Langmuir probe and an emissive probe. More details about the complete evolution of the plasma parameters over a broad range of discharge conditions can be found in Ref. \cite{surbhirsi}.   \par
Melamine Formaldehyde (MF) mono-dispersive spherical particles of different sizes (10.66, 8.90 and 4.38 $\mu$m) are then introduced into the plasma by a dispenser to create a dusty plasma. These micro-particles collect more electrons than ions from the plasma due to the higher mobility of electrons and get negatively charged and levitate near the cathode sheath region due to a balance between the electrostatic force (acting in the upward direction) and the gravitational force (acting in the downward direction). To observe the dynamics of these dust particles, they are illuminated by a green laser light (in the $x-z$ plane). The Mie-scattered light from the dust particles are captured by a CCD camera and the images from the camera are stored into a computer for further analysis. \par 
 \section{Flow generation and velocity Measurement} \label{sec:flow}
For conducting flowing dusty plasma experiments, these MF particles are made to flow along the axis of the $\Pi$-tube using two different methods e.g., Single Gas Injection (SGI) and Dual Gas Injection (DGI) techniques as discussed in details in \cite{surbhiflow}. In the SGI technique, a steady state dust cloud is initially formed in between the two axial confining strips by precisely balancing the pumping rate and the gas flow rate. This equilibrium is then disturbed momentarily either by increasing the pumping rate or by reducing the gas flow using a mass flow controller (controlled by a software through a computer) attached to port-P$_1$. In both the cases the particles are seen to move from the right to the left along the axis of the chamber. For our present set of experiments, the flow in dust particles was always initiated by reducing the gas flow in steps of $\sim 0.5$ $sccm$. In the DGI method of flow generation an equilibrium working pressure is maintained by introducing Argon gas using both the gas ports P$_1$ and P$_2$ as shown in Fig.~\ref{fig:fig1}. The dust particles are then injected into the plasma with the help of a dust dispenser and they get charged during the time of their fall. These particles come to the sheath region and are seen to flow from right to left due to the continuous flow of neutral gas. It is worth mentioning that for carrying out all these experiments the pressure range as well as the flow rate differences are set in such a way that the flow of dust particles is always laminar in nature. The dynamics of dust particles for these two different flow generation techniques will be discussed in subsequent sections. \par
\begin{figure}[ht]
\centering{\includegraphics[scale=0.3]{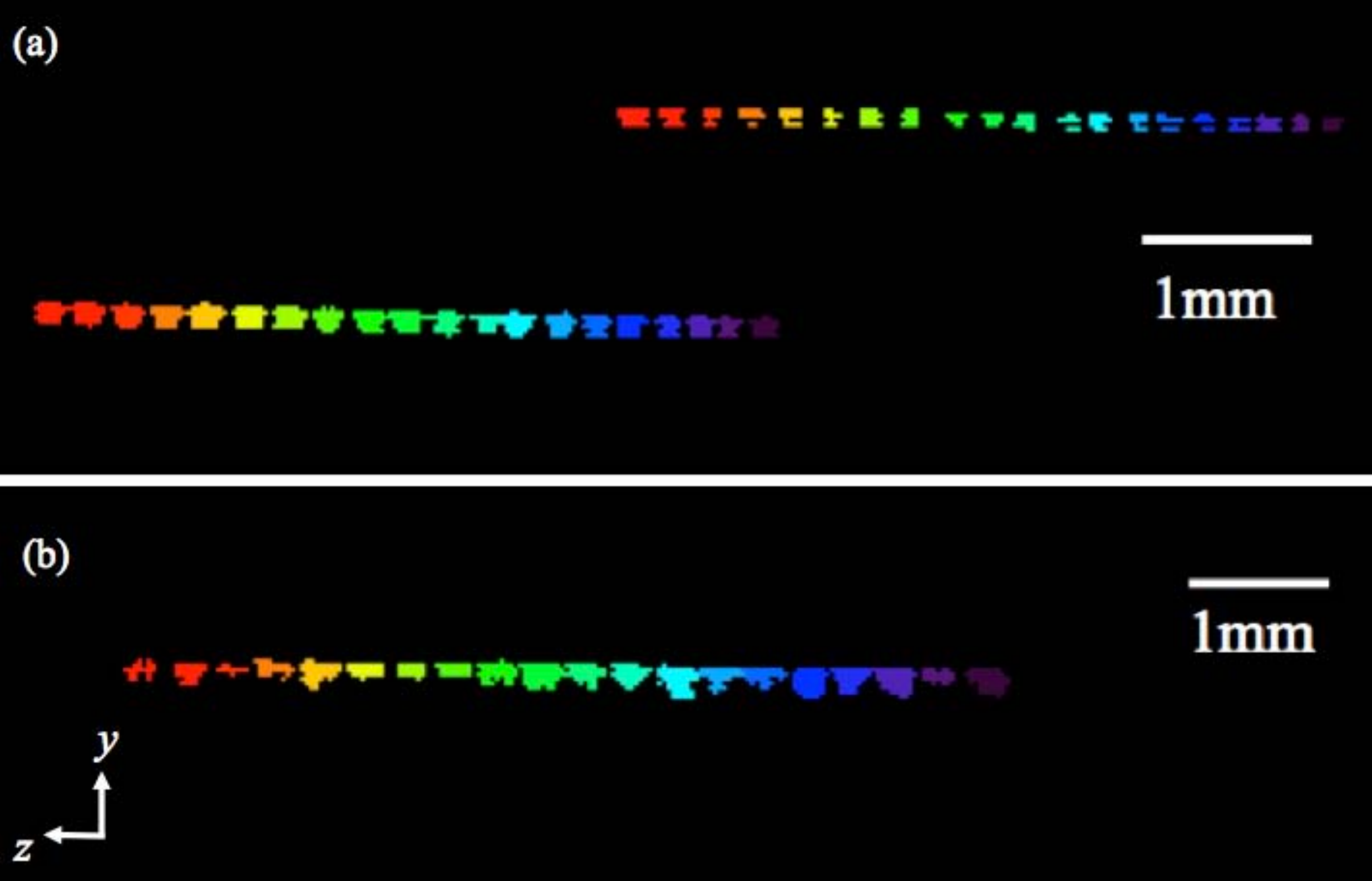}}
\caption{\label{fig:fig2} Color plot showing the trajectories of two particles where the flow is generated using (a) Single Gas Injection Technique (SGI) (b) Dual Gas Injection Technique (DGI).}
\end{figure}
Fig.~\ref{fig:fig2}(a) shows the trajectories of two dust particles of diameter 10.66 $\mu$m at a given pressure of 12 Pa and a discharge voltage of 320 V when the flow is generated using the SGI technique. Different colors represent the particle positions at different times in intervals of 6.25 msec. Violet color corresponds to initial position of the particle whereas the red color corresponds to final position of the particle. The figure shows that the particle moves from right to left along the axial direction (along $z$) almost with a constant velocity.  Fig.~\ref{fig:fig2}(b) shows the same when the flow is initiated using the DGI technique. It is to be noted that the trajectories show the same characteristics when the particles attain a constant velocity in both the techniques.\par 
To study the dynamics of these dust particles, it is very important to measure the axial and radial components of the velocity. It essentially helps us to understand the fundamental forces that act on them. For analysing the video images of the flow of these dust particles an Idl based super Particle Identification Tracking (sPIT) code \cite{konopka2000wechselwirkungen,Feng} is used in which hundreds of still frames are considered over a wide range of discharge conditions. This code is very powerful and efficient to track individual particles and to measure the velocity very precisely when the particles are well resolved during the span of their journey. 
\section{Results and Discussion}
\label{sec:results}
\subsection{Influence of gas flow rate and neutral gas pressure on the dust particle dynamics}\label{sec:results1}
\begin{figure}[ht]
\centering{\includegraphics[scale=1.0]{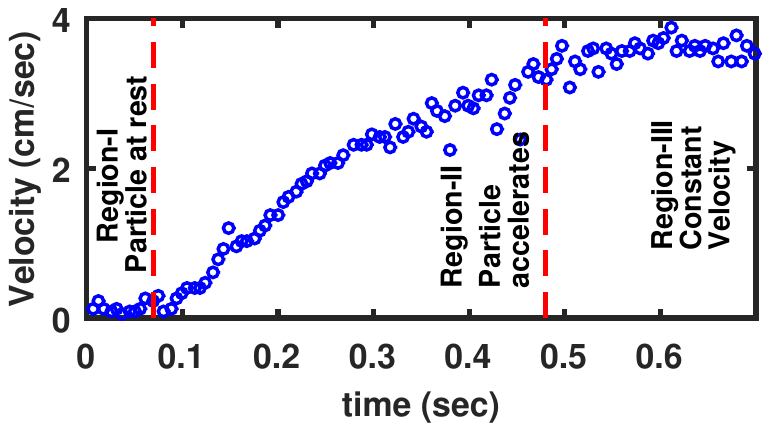}}
\caption{\label{fig:fig3}Time evolution of 10.66 $\mu$m particle at a fixed pressure of p=12 Pa and discharge voltage of 300 V when the flow is generated by Single Gas Injection Technique (SGI).}
\end{figure}

To begin with, a series of experiments were carried out to study the dynamics of the dust particles by changing the gas flow rate and the background pressure. Fig.~\ref{fig:fig3} shows the time evolution of a dust particle having diameter 10.66 $\mu$m when the experiments were performed at a discharge voltage of 300~V and pressure of 12~Pa and in which the flow was initiated by the SGI technique \cite{surbhiflow} with a flow rate difference of 3.4~$sccm$ . 
To understand the dynamics of dust particle in a better way, Fig.~\ref{fig:fig3} is divided into three different distinct regions. In region-I (region of zero velocity), the dust particles only exhibit random motion (no directional motion) and represent the phase when no flow in the dust particles has yet been initiated. Region-II is the accelerating region, where the dust particles start their journey from rest in the $z$-direction and accelerate for approximately 0.4 seconds. This happens when the gas flow rate near the port P$_1$ is suddenly reduced and as a result the neutrals move from right to left for neutralizing the imbalance of sudden gas pressure. On their way towards the pump, the neutral molecules impart their momentum to the heavier dust particles causing them to move in the direction of the neutrals. The transfer of momentum from neutrals to dust particles continues till the velocity of the dust particles become equal to the velocity of the neutrals and correspond to the accelerating phase marked as Region-II in Fig.~\ref{fig:fig3}. When the velocity of the dust particle becomes equal to the velocity of neutrals, it moves with almost constant velocity - an asymptotic velocity as depicted in region-III. In this region of Fig.~\ref{fig:fig3}, it is clearly seen that the particle moves with a constant velocity of $\sim 4$~cm/sec during $~0.6 -0.8$~sec. It is worth mentioning that the moving neutrals impart their momentum to dust particle whenever the velocity of the dust particle decreases due to other damping phenomena.   \par    
\begin{figure}[ht]
\centering{\includegraphics[scale=0.85]{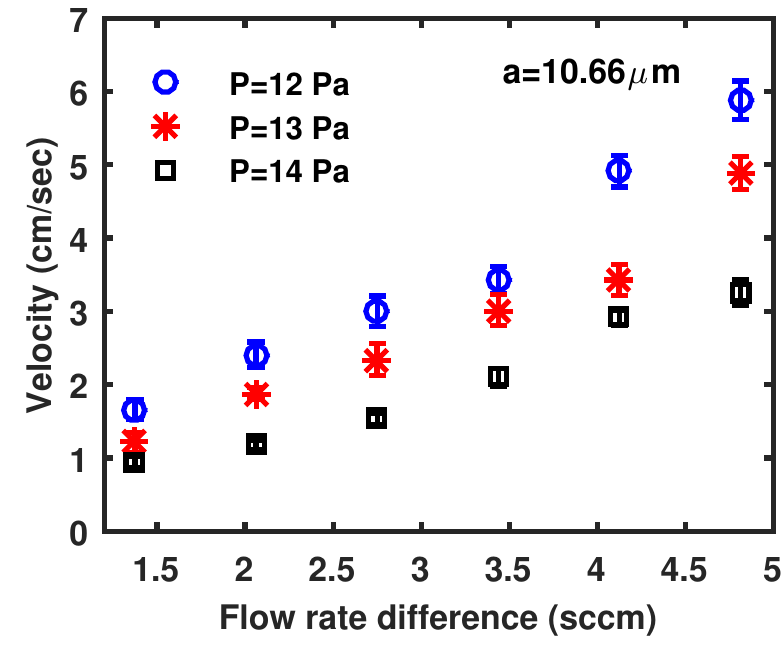}}
\caption{\label{fig:fig4} Variation of asymptotic velocity of 10.66 $\mu$m particle with flow rate difference for three different gas pressures.} 
\end{figure}
Fig.~\ref{fig:fig4} shows the variation of this asymptotic velocity achieved by 10.66 $\mu$m particle when the differential gas flow rate is varied by different amounts and for three different background pressures. The average asymptotic velocities are calculated from the time evolution of different particles with the same size at the time when they attain a constant velocity as discussed in Fig.~\ref{fig:fig3}. The figure indicates that the asymptotic velocity increases monotonically with the gas flow rate difference. A higher flow rate difference concomitant larger transfer of momentum to the dust particles which then moves with a greater velocity for a given pressure. Fig.~\ref{fig:fig4} also shows that the asymptotic velocity decreases with an increase of the background gas pressure for a constant flow rate difference. This can be understood from the fact that the number of neutral gas molecules increases with the increase of Argon gas pressure and as a result the neutrals undergo frequent collisions among themselves which essentially reduces the effective directional velocity of the neutrals. As a consequence they impart less directional momentum to the dust particles causing them to move with a smaller velocity.\par
\subsection{Dynamics of particles with different sizes}\label{sec:results2}
\begin{figure}[ht]
\centering{\includegraphics[scale=0.9]{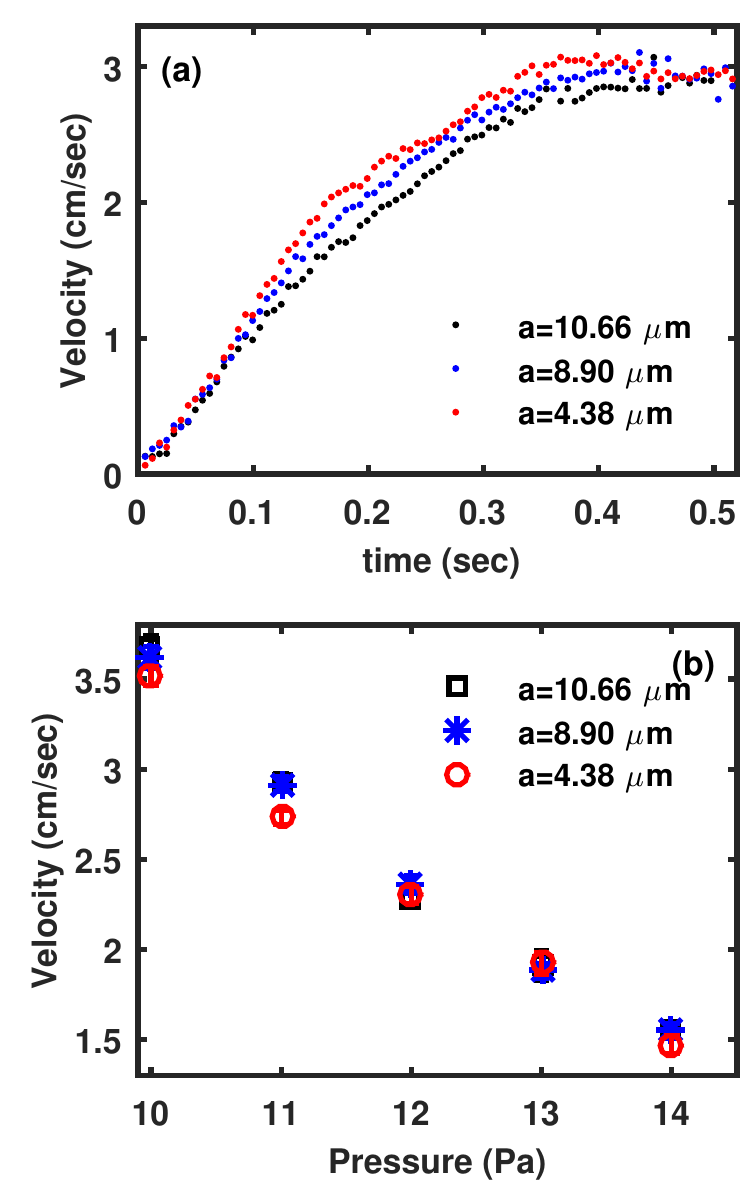}}
\caption{\label{fig:fig5} Variation of constant velocity with pressure for three different micro particles of diameter 10.66 $\mu$m, 8.90 $\mu$m and 4.38  $\mu$m when the flow is generated by SGI technique}    
\end{figure}

To study the effect of particle sizes on the flow velocity, a set of experiments was carried out with different sized dust particles. Fig.~\ref{fig:fig5}(a) shows the trajectories of dust particles having diameter 10.90 $\mu$m, 8.90 $\mu$m and 4.38 $\mu$m for the case of gas flow rate difference of $\sim$ 2.1 $sccm$ at a constant pressure of 11 Pa and discharge voltage of 300 V.  Fig.~\ref{fig:fig5}(a) shows that the asymptotic velocity attained by all these micron sized particles are nearly equal indicating that the Region-III as shown in Fig.~\ref{fig:fig3} is independent of the sizes and hence masses of these dust particles. However, the acceleration region (Region-II) shows a marked dependency on the size of the dust particle. The trajectories show that the lighter dust particle initially moves with a higher velocity than the heavier one whereas after $\sim 0.4$ sec, the lighter particle achieves the same constant velocity as attained by the bigger particle. In the accelerating region in which the dust particle gains momentum from the moving neutrals, the acceleration is higher for the lighter dust particle whereas it is smaller for the heavier particle as shown in Fig.~\ref{fig:fig5}(a). \par
Finally, the asymptotic velocities for these three different sized particles were measured over a range of pressures from 10 Pa to 14 Pa by keeping the other discharge parameters constant at a flow rate difference of $\sim$ 2.1 $sccm$ and the results are shown in Fig.~\ref{fig:fig5}(a).  Fig.~\ref{fig:fig5}(b) shows the variation of the asymptotic velocities of these three different sized particles with pressure. It is clearly seen that all the particles attain the same velocity for the entire pressure range and this velocity decreases with an increase in the background pressure. The error bars shown in the figure are calculated by performing the experiments several times and taking the mean as well as standard deviation of velocities.\par
 To examine the dynamics of dust particles in a steady state equilibrium, another set of experiments was carried out by introducing  Argon gas through the gas port P$_2$ with the help of Dual Gas Injection technique as described in section \ref{sec:flow} and in Ref. \cite{surbhiflow}.  It should be mentioned that, in this technique of flow generation it is difficult to identify Region-I (region of zero velocity) and Region-II (region of acceleration) of the particle trajectory as identified in Fig.~\ref{fig:fig3} for the SGI technique as the particles attain their asymptotic velocity nearly from the very beginning of their journey. It is due to the fact that there is a continuous flow of neutrals from port $P_2$ towards the pump which always carry the dust particles as soon as they fall into the sheath region. \par
\begin{figure}[ht]
\centering{\includegraphics[scale=0.8]{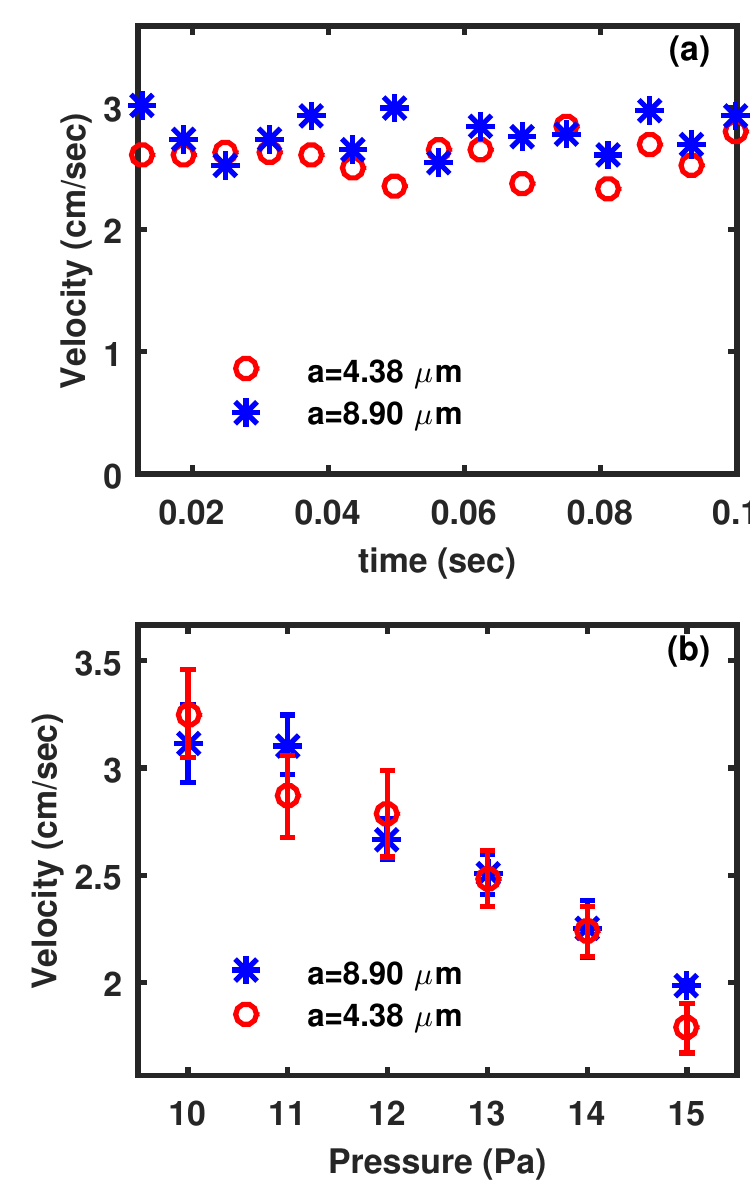}}
\caption{\label{fig:fig6}(a) Time evolution and (b) variation of constant velocity with pressure of 8.90 and 4.38 $\mu$m micro particle when flow is generated by Dual Gas Injection (DGI) technique.}
\end{figure}
 Fig.~\ref{fig:fig6}(a) shows the time evolution of the velocity for two different dust particles of diameter 8.90 $\mu$m and 4.38 $\mu$m, respectively when the flow in the dust particle is generated by the DGI technique at a discharge voltage of 300 V and pressure of $p=13$ Pa. It is seen from the figure that irrespective of  sizes, the particles move at constant velocities of nearly the same value. Fig.~\ref{fig:fig6}(b) shows that the asymptotic velocity decreases monotonically with increasing background pressure akin to the earlier case (see Fig.~\ref{fig:fig5}(b)) when the flow is generated using SGI technique. Interestingly, it is also seen that both the particles possess nearly the same velocity for the entire range of gas pressure. Therefore, the experimental results obtained in both the techniques demonstrate that the neutrals are primarily responsible for carrying the dust particles in their own direction of flow from the right to the left. We will establish a more quantitative confirmation of this observation in Sec.~\ref{sec:forces} by making theoretical comparisons between all other forces acting on the dust particle and comparing them with the neutral streaming force. \\   
 \subsection{Influence of nature of background gas on the dynamics of dust particles}\label{sec:results3}
\begin{figure}[ht]
\centering{\includegraphics[scale=0.8]{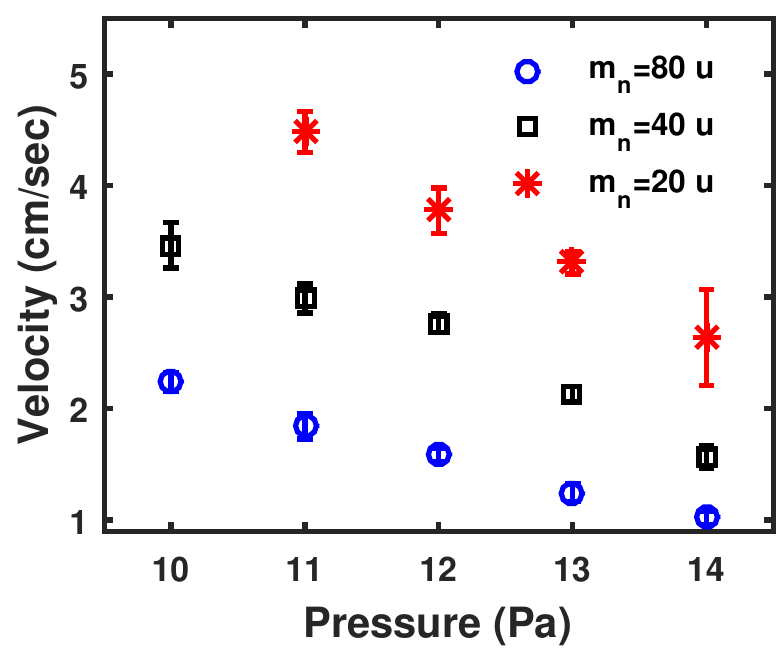}}
\caption{\label{fig:fig8}Variation of constant velocity with pressure for Krypton, Argon, Neon gases. }
\end{figure}

For studying the role of different neutral species on the dynamics of dust particles, an experimental investigation was carried out using three different gases with different masses.  The average asymptotic velocities were measured from the trajectories of the particles when they attained their constant velocities. In this set of experiments the flow rate difference and other discharge parameters were kept constant at 2.75~$sccm$, $300-500$ V and $10-15$ Pa, respectively. The variation of these asymptotic velocities for different gases are plotted in Fig.~\ref{fig:fig8}. 
The open-circle, square, and star represent the results of the dust particle velocities in backgrounds of Krypton ($m_n=80$ $u$), Argon ($m_n=40$ $u$), and Neon ($m_n=20$ $u$) gases, respectively. It is observed that the asymptotic velocities again decrease with an increase of gas pressure for all the gases as seen before in Fig.~\ref{fig:fig5}(b) and Fig.~\ref{fig:fig6}(b). However, for a given pressure the dust particles move with a slower asymptotic velocity (represented by open circles) in a background of heavier neutrals compared to a background of lighter gas molecules (represented by open stars).  This can be understood from the fact that for a given flow rate difference the lighter neutrals move with a higher velocity due their lower inertia compared to the heavier neutrals. Since the asymptotic velocity of the dust particle equals the flow velocity of the neutrals they move faster in a gas of lighter mass.  As a result the dust particles move with maximum velocity in the case of Neon plasma, whereas they move with minimum velocity for Krypton plasma. \\
 \subsection{Estimation of forces acting on the dust particles:}\label{sec:forces}
 For an estimate of the various forces (e.g. neutral streaming force, ion drag force, electrostatic force \textit{etc.}) acting on the particle, first the electric fields ($E=-\frac{dV_p}{dZ}$) are estimated over a wide range of discharge parameters by measuring the plasma potential ($V_p$) at different axial locations. For that estimation, the average plasma potentials ($V_p$) are measured (by floating point technique \cite{sheehan}) using a couple of emissive probes that are placed at two different axial locations at a distance of $dZ=5$ cm.  Finally, the effective electric field is  estimated from the difference of plasma potentials before and during the generation of flow in dust particles. It is found that during the flow generation the maximum electric field turns out to be in the range of $8-12$ V/m for a flow rate difference of $1.5-5$ $sccm$, $V_d=300$~ V and $p=10-15$~Pa. \par
 \subsubsection{The Coulomb force:}
To start with, the Coulomb force ($F_e=Q_dE$) acting on the dust particles is determined by multiplying the charge ($Q_d$) acquired by the dust particles with the local axial electric field as discussed before. The charge is calculated by assuming the dust particle to be a spherical capacitor and using the fundamental relation, $Q_d = CV_s$, where $C=4\pi\epsilon_0 r_d$ and $V_s$ are capacitance and surface potential of dust particle. The surface poatential of dust particle is calculated using Collision Enchanced plasma Collection model (CEC) \cite{khrapak2,ratynskaia2004} and $Q_d$ comes out to be $\sim 2.87\times10^{-15}$ C of $r_d=5.33$ $\mu$m particle. For our experimental parameters, $p=12$ Pa, flow rate difference $3.5$ $sccm$, $E=10$ V/m and $T_e=3.0$~eV which yields the value of Coulomb force, $F_e\sim 30$~fN.
  \subsubsection{Ion drag force:}
  For the ion drag force $F_i$, the expression given by Khrapak et al. \cite{khrapak2005hybrid} is used, which estimates $F_i$ for a single micro-particle in a collisionless Maxwellian plasma for an arbitrary ion velocity as:
 \begin{equation*}
\begin{aligned}
F_i = 
& \sqrt{2\pi}r_d^2n_im_iv_{ti}^2 \biggr\{ \sqrt{\frac{\pi}{2}} \text{erf}\left (\frac{u_f}{\sqrt2}\right) \times\\
\bigl[ & 1+ u_f + (1-u_f^{-2})(1+2z\tau)+4z^2\tau^2u_f^{-2}\text{ln}\Lambda\bigl] \\
+ & u_f^{-1} \bigl[ 1+2z\tau+u_f^{2}-4z^2\tau^2\text{ln}\Lambda\bigl] \text{exp}\left(-\frac{u_f^2}{2}\right) \biggr\}.
\end{aligned}
\label{eqn:ion_drag}
\end{equation*}
where, $n_i$, $m_i$,  $v_{ti}=\sqrt{8K_BT_i/m_i\pi}$ are the density, mass and thermal velocity of ions, respectively. The ion drift velocity normalized by the ion thermal velocity is denoted by $u_f$. The dimensionless charge of the micro particle is given by $z = Q_d|e| /(4\pi r_d K_B T_e)$, and the ratio of the electron to the ion temperature is denoted by $\tau = T_e/ T_i$. The Coulomb logarithm is given by:
\begin{equation}
\text{ln}\Lambda = \text{ln}\left[\frac{\beta+1}{\beta+(r_d/\lambda_{ef})}\right]\nonumber
\end{equation}
where $\beta$ is defined as $\beta = Q_d|e|/[4\pi\epsilon_0k_BT_i(1+u_f^2)\lambda_{ef}]$ and $\lambda_{ef}(u_f) = 1/\sqrt{\lambda_i^{-2}(1+u_f^2)^{-1}+\lambda_e^{-2}}$ is the effective screening length. The ion flow velocity is calculated by using the formula $v_i = \mu_i E$. E is the effective electric field when the flow in the dust particle is initiated and $\mu_i$ is the ion mobility, estimated by using the expression $\mu_i(E) = \mu_0/p\sqrt{(1+\alpha E/p)}$ with $\mu_0 = 19.5~ $m$^2$PaV$^{-1}$s$^{-1}$, $\alpha = 0.035~$mPaV$^{-1}$ for Argon gas of pressure p (in Pascal) \cite{schwabe2017observation}. For E=10V/m,  p=12 Pa, $n_i=1.2\times10^{15}$ m$^{-3}$, the ion drag force comes out to be $F_i\sim 9$~fN. 
\subsubsection{Neutral streaming force:}
The force on the dust due to streaming neutrals can be calculated by the Epstein formula \cite{epstein}, as expressed by 
\begin{equation*}
F_{n}=\frac{4}{3}\delta{\pi}r_d^2m_nn_nv_{tn}(v_f-v_d),
\label{eqn:fn}
\end{equation*}
where, $m_n$, $n_n$, $v_{tn}$ and $v_f$ are the mass, number density, thermal and drift velocities of the neutrals, respectively. It is worth mentioning that the neutral streaming force acting on the dust particles is proportional to the relative velocity of neutral with respect to dust. It becomes maximum when the dust particles are at rest and minimum when they attain the velocity of neutrals. $\delta$ represents the Epstein drag coefficient which varies from $1$ to $1.4$ depending upon the types of reflection \cite{epstein}. The Epstein coefficient as measured in our device is $\delta \sim 1.2$ \cite{surbhirsi}. For a specific set of experiments, the neutral velocity comes out to be $v_f=4$~cm/sec at the flow rate difference of 3.50 $sccm$ and background Argon gas pressure of $p=12$ Pa. The particles of diameter 10.66 $\mu$m, $m_n = 6.67 \times 10^{-26}$ kg, $n_n\sim 3\times 10^{21}$ m$^{-3}$ and thermal velocity $v_{tn}\sim 430$~m/sec at a temperature of 300 K, the magnitude of maximum neutral streaming force comes out to be $\sim$ 700 fN. The above quantitative estimates show that the neutral streaming force is the predominant one for our experimental conditions and is primarily responsible for making the dust particles flow.
\subsection{Theoretical model}
\label{sec:model}
\par To interpret the experimental findings qualitatively, a simple theoretical model is developed in which the equation of motion for the dust particle can be written as:
\begin{equation}
m_d \frac{dv_d}{dt}=\delta \frac{4}{3} \pi r_d^2m_nn_nv_{tn}(v_f-v_d).
\label{eqn:EM}
\end{equation} 
$\delta$ is the Epstein drag coefficients \cite{surbhirsi,liu2003radiation} and $m_d$ is the mass of MF dust particle of diameter 10.66 $\mu$m. It is to be noted that only the contribution of neutral streaming force is taken into consideration in the right hand side of Eq.~\ref{eqn:EM} as the other forces are insignificant compared to the neutral streaming force as discussed before in subsec.~ \ref{sec:forces}. Eq.~\ref{eqn:EM} also reveals that at $v_f=v_d$, the total force acting on the particle becomes zero and as result the particle moves with a constant velocity. Rewriting v=$v_d/v_f$, Eq.~\ref{eqn:EM} simplifies as:
\begin{equation}
m_d \frac{dv}{dt}+cv=c,
\label{eqn:sEM}
\end{equation} 
where $c=\delta\frac{4}{3} \pi r_d^2m_nn_nv_{tn}$.\par
After solving  Eq.~\ref{eqn:sEM} we get:
\begin{equation}
v_d=v_f(1-e^{-ct/m_d}).
\label{eqn:sol}
\end{equation}
\begin{figure}[ht]
\centering{\includegraphics[scale=0.8]{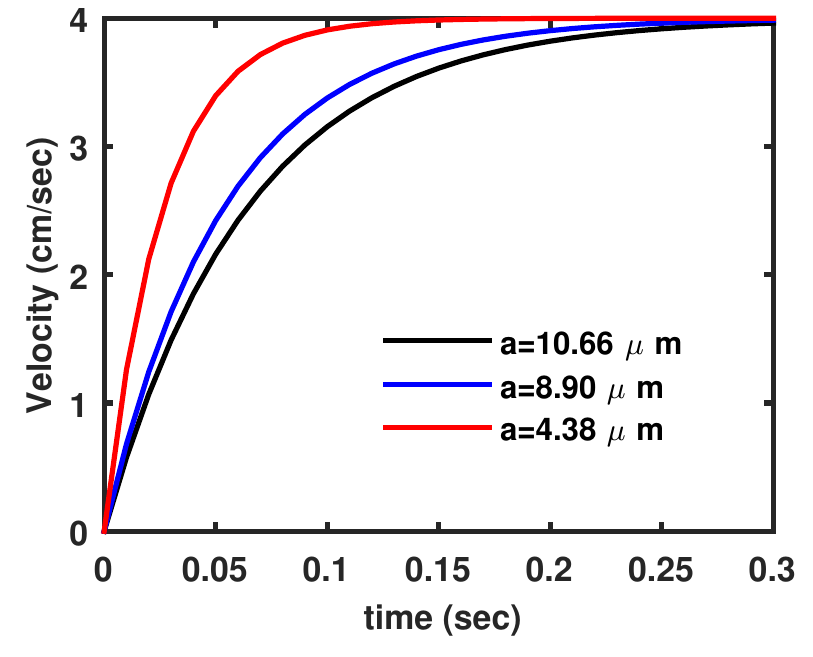}}
\caption{\label{fig:fig7} Theoretical plot of time evolution of 10.66, 8.90 and 4.38 $\mu$m particles.}
\end{figure}
The solution (Eq.~\ref{eqn:sol}) of equation of motion (Eq.~\ref{eqn:EM}) essentially describes the complete dynamics of the dust particle under the influence of the neutral streaming force. From (\ref{eqn:sol}) it is seen that at $t=0$ the particle starts from rest ($v_d=0$) and asymptotes ($t\to\infty$) to $v_f$ the flow velocity of the neutrals. \par 
For our experimental value of the neutral velocity, $v_f = 4\; cm/s$, we use the theoretical equation (\ref{eqn:sol}) to plot the time evolution of the velocities for three different sizes of dust particles in Fig.~\ref{fig:fig7}. As can be seen, all the particles start from rest but the lighter particles gain velocity rapidly but they all asymptote to the same final value in agreement with the experimental results shown in Fig.~\ref{fig:fig5}.
\section{Conclusions}
\label{sec:conclusions}
In conclusion, a set of experiments has been carried out to investigate the role of neutrals in initiating and sustaining the flow of dust particles in the Dusty Plasma Experimental (DPEx) device. The characteristics of the dust flow, initiated by two different techniques, namely the Single Gas Injection technique (SGI) and the Dual Gas Injection Technique (DGI), have been studied as a function of the neutral gas flow, the background pressure and the nature of the background gas. Our principal findings are as below: 
\begin{enumerate}
\item{In the case of the SGI technique, the dust particles start from rest, go through an accelerating phase and then attain a steady state constant velocity. In the DGI method the acceleration phase is not discernible experimentally and the dust particles are carried along continuously by the flow of neutrals at a constant velocity.}
\item{The asymptotic steady state velocity increases with an increase of the gas flow rate and decreases with the increase of neutral background pressure when the gas flow rate is kept constant for both the techniques.}
\item{The asymptotic velocity is independent of the size of the dust particles but in the accelerating phase (for the SGI technique) the lighter dust particles show a higher acceleration compared to the heavier ones.}
\item{From experiments carried out with different background gases it is also found that the asymptotic velocity is dependent on the mass of the neutral molecules. The velocity is higher in the presence of a lighter gas compared to a heavier gas.} 
\item{The experimental results can be well understood in terms of the neutral streaming force acting on the dust particles.}
\item{A quantitative estimate of the various forces acting on the dust particles for our experimental conditions indicates that the neutral drag force predominates over all other forces and is the primary driver of the dust flow. A simple analytical solution of the equation of motion of the dust particle under the influence of the neutral drag force reproduces the principal features of the experimental results.}
\end{enumerate}
Our results thus provide a detailed picture of the micro-dynamics of dust flow initiated and utilized in past 
experiments on the DPEx device for the excitation of various linear and nonlinear coherent structures and should also prove useful in gaining understanding of dust flow related phenomena in the laboratory or in nature.\\\\
\textbf{References}\\
\providecommand{\newblock}{}

\end{document}